\documentstyle[twoside,fleqn,npb,epsfig]{article}
\newcommand{\bm}[1]{\mbox{\boldmath $#1$}}
\newcommand{\be}{\begin{equation}}
\newcommand{\ee}{\end{equation}}
\newcommand{\bea}{\begin{eqnarray}}
\newcommand{\eea}{\end{eqnarray}}
\newcommand{\st}{{\scriptscriptstyle T}}

\def\slash{\rlap{/}}

\newcommand{\AmS}{{\protect\the\textfont2
  A\kern-.1667em\lower.5ex\hbox{M}\kern-.125emS}}

\title{Estimates of T-odd distribution and fragmentation functions}

\author{M. Boglione and P.J. Mulders \address{ 
             Vrije Universiteit Amsterdam,
            De Boelelaan 1081, 1081 HV Amsterdam, The Netherlands}
}

\begin{document}

\begin{abstract}
Estimates of the T-odd fragmentation and distribution functions, 
$H_1^\perp$ and $f_{1T}^{\perp}$, are presented. Our evaluations are 
based on a fit on experimental data in p$^\uparrow$p. 
We use our estimates to make predictions for ep$^\uparrow$ azimuthal 
asymmetries.
\end{abstract}

\maketitle

Distribution and fragmentation functions account for the soft parts of a 
scattering process in which quarks are produced from the initial hadrons, 
and final hadrons are produced from quarks resulting from the elementary 
hard scattering.
Leading order distribution and fragmentation functions have a direct 
interpretation in terms of probability densities (see Ref.~\cite{bm99} 
for more details and pictures).

In this talk, we focus our attention on the distribution and the fragmentation 
functions $f_{1T}^{\perp a}(x)$ and  $H_1^{\perp}(z)$, which  are T-odd 
functions, i.e. they are not constrained by time reversal invariance. 
The function $H_1^{\perp}(z)$, for which the non applicability of time 
reversal symmetry is straightforward,  
allows for processes in which transversely polarized quarks fragment into 
an unpolarized hadron. 
In the less straightforward situation where time reversal symmetry cannot be 
applied  for distribution functions 
~\cite{abm95,adm96,gluonicpoles}, a non-zero $f_{1T}^{\perp a}(x)$  allows for 
processes in which unpolarized quarks are produced from a polarized proton.
 
Our estimates are based on
the parametrizations presented in Ref.~\cite{abm95,am98,abm98}, 
obtained from fits on the FNAL E704 experimental data 
on single spin asymmetry in $p^{\uparrow}\,p \,\to \, \pi \,X$.
These allow us to evaluate some weighted integrals, proposed in 
Ref.~\cite{bm98}, which are directly related to measurable physical 
observables, the angle $\phi ^l _h$ between the 
lepton scattering plane and the produced hadron plane,
and  the angle $\phi ^l _S$  between 
the lepton scattering plane and the nucleon spin.
Finally, we evaluate the ratio $H_1^{\perp}/D_1$ 
and $f_{1T}^{\perp}/f_1$ and compare them with existing experimental data.


Applying Lorentz invariance, hermiticity, and parity invariance 
to the general lightfront correlator~\cite{correlation}, one finds that, 
as far as relevant at leading order in $1/Q$, its Dirac structure is given 
by (see Ref.~\cite{bm98} for details)
\bea
\lefteqn{\Phi(x,\bm k_\st; P,S) = \frac{1}{4}\Biggl\{
f_1\,\slash \slash n_+
+ h_1^\perp\,\frac{\sigma_{\mu \nu} k_\st^\mu n_+^\nu}{M}
} \nonumber \\ &&
+ f_{1T}^\perp\, \frac{\epsilon_{\mu \nu \rho \sigma}
\gamma^\mu n_+^\nu k_\st^\rho S_\st^\sigma}{M} 
+h_{1T}\,i\sigma_{\mu\nu}\gamma_5 n_+^\mu S_\st^\nu
\nonumber \\ &&
+g_{1s}\, \gamma_5\slash n_+
+h_{1s}^\perp\,\frac{i\sigma_{\mu\nu}\gamma_5
n_+^\mu k_\st^\nu}{M}
\Biggr\}.
\label{param}
\eea
Just as for the distribution functions,
the full Dirac structure relevant for fragmentation
into spin 0 (or unpolarized) 
hadrons, up to leading order, is given by
\be
\Delta(z,\bm k_\st,P_h) =
\frac{1}{2}\Biggl\{
D_1\,\slash \slash n_-
+ H_1^\perp\,\frac{\sigma_{\mu \nu} k_\st^\mu n_-^\nu}{M_h}
\Biggr\}.
\label{DDelta}
\ee
The link with the helicity formalism, used in Refs.~\cite{abm95,abm98}, 
is achieved by transforming the $\Phi_{ij}$ matrix elements to the 
helicity basis through the density matrix $\rho$
\be
\Phi_{ij} (x,k_\st;P,S) = \sum_{\Lambda \Lambda ^{\prime}}
\rho _{\Lambda \Lambda ^{\prime}} (S)
\; \Phi _{\Lambda i;\, \Lambda ^{\prime} j} (x,k_\st;P),
\ee
where $\Lambda$, $\Lambda ^{\prime}$ are the helicity indices of the proton
and $S$ the spin vector, and $\rho _{\Lambda \Lambda ^{\prime}}$ is defined as
\be
\rho _{\Lambda \Lambda ^{\prime}} = \frac{1}{2} \;
( \delta _{\Lambda\Lambda^\prime}
+ \bm S \cdot (\bm \sigma)_{\Lambda \Lambda^\prime}) \; .
\ee
In the rest-frame, where $S = (0,\bm S_\st,\lambda)$, one obtains
\bea
\lefteqn{ \Phi_{ij}(x,\bm k_\st, P,S) = 
\frac{1}{2}\Bigl(\Phi_{+i;\,+j} + \Phi_{-i;\,-j}\Bigr) } \nonumber \\ &&
\hspace{-1.1cm}
+\frac{1}{2}\,S_\st^1\Bigl(\Phi_{+i;\,-j} +\Phi_{-i;\,+j}\Bigr)
-\frac{i}{2}\,S_\st^2 \Bigl(\Phi_{+i;\,-j} - \Phi_{-i;\,+j}\Bigr)
\nonumber \\ &&
+\frac{1}{2}\,\lambda\Bigl(\Phi_{+i;\,+j} - \Phi_{-i;\,-j}\Bigr)\,.
\label{helicity}
\eea
By comparing Eqs. (\ref{param}) and (\ref{helicity}), 
term by term, one can see that the term proportional to
$f_{1T}^{\perp}$ in the $\Phi _{ij}^{[\gamma ^{+}]}$ projection can be
identified with the function $\Delta^Nf_{q/\uparrow}$ = $2\,I_{+-}$ 
defined in Ref.~\cite{abm95}.
To be more precise, one finds
\be
\Delta^Nf_{q/\uparrow}(x) 
= 2\,
\frac{\langle  k_\st (x) \rangle}{M} \; f_{1T}^{\perp}(x,\bm k_\st).
\ee
In later applications it will turn out to be useful to consider the
$(\bm k_\st^2/2M^2)$ weighted function
\be
f_{1T}^{\perp (1)}(x) = \int d^2k_\st\ \frac{\vert \bm 
k_\st\vert^2}{2M^2}\,f_{1T}^\perp (x,\bm k_\st)\;,
\label{wf1Tperp}
\ee
for which we use the estimate
\be
f_{1T}^{\perp (1)}(x)  
= \frac{\langle k_\st(x)\rangle}{4M}\,\Delta^Nf_{q/\uparrow}
(x)\;.
\label{f(1)}
\ee
Using the results from the
most recent analysis of the pion left-right asymmetry in 
p$^\uparrow$p $\rightarrow \pi$X in Ref.~\cite{am98} (see also footnote 
in~\cite{abm98}),
and the results from, for example, Ref.~\cite{jrr89} for the average
transverse momentum,$\langle k_\st(x)\rangle$,
we obtain for
$f_{1T}^{\perp (1)}$ the estimate
\bea
f_{1T}^{\perp (1)\,u}(x) &=& 0.81\;x^{2.70}\;(1-x)^{4.54}\; , \nonumber \\
f_{1T}^{\perp (1)\,d}(x) &=& -0.27\;x^{2.12}\;(1-x)^{5.10} .
\label{f1}
\eea
Similarly, for the fragmentation function $H_1^\perp$ we find
\be
\Delta ^N D (z,\bm k_\st) = 
-2\,\frac{\langle \bm k_\st (z)\rangle}{M_h}
\,H_1^{\perp}(z,\bm k_\st^\prime)\;,
\label{Delta}
\ee
and
\be
H_1^{\perp (1)}(z) = -\frac{\langle k_\st(z)\rangle}{4M_h}
\,\Delta ^N D(z).
\ee
Making use of the results of Ref.~\cite{abm98},
and of a fit to the
LEP data~\cite{abreu},
we find
\be
H_{1}^{\perp (1)}(z) = 1.08\;z^{2.87}\;(1-z)^{0.64}\;.
\label{H(1)}
\ee
\begin{figure}[t]
\vspace{-1cm}
\mbox{~\epsfig{file=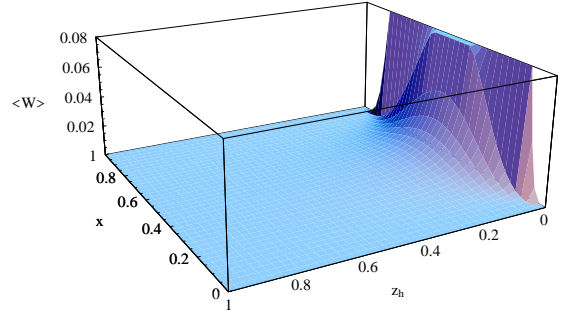,angle=0,width=7.5cm}}
\vspace{-1cm}
\hspace{-0.4cm}
\caption{A three-dimensional view of the quantity 
$\sum _{a,\overline a} e_a ^2 x \, 
f_{1T}^{\perp (1) a}(x)  \, H_1 ^{\perp (1) a} (z_h)$, 
for scattering of unpolarized leptons on a polarized proton target, 
with production of $\pi ^+$}
\label{W2}
\vspace{-0.3cm}
\end{figure}
We now have all the ingredients to calculate the weighted integrals proposed 
in Ref.~\cite{bm98}. Following the notations introduced therein,
we will focus our attention on the following two of such objects.

\bea
\lefteqn{\left< \frac{Q_T}{M}\;\sin(\phi^l_h - \phi^l_S) \right>_{OTO} = }
\nonumber \\ & & \hspace{-.7cm}
\frac{ 4 \pi \alpha ^2 s}{Q^4} \, (1-y)\,
\sum _{a,\overline a} e_a ^2 \ x \, f_{1T}^{\perp (1) a}(x) \, D_1 ^a(z_h)\;.
\label{f1D}
\eea

A~three-dimensional~plot~of~the~quantity \\ $\sum _{a,\overline a} 
e_a ^2 \ x \, f_{1T}^{\perp (1) a}(x) \, D_1 ^a(z_h)$ is shown in 
Fig.~\ref{W2}. 
The shape of the surface as a function of $x$ and 
$z_h$ tells us that the effect due to the T-odd distribution function becomes 
sizeable for very small values 
of $z_h$ and intermediate values of $x$. 
It is clear that the effects due to 
the presence of the T-odd  distribution function $f_{1T}^{\perp}(x)$ are 
small, but a suitably designed experiment may put limits on their size, 
or might establish their mere existence. This would be a crucial test for 
the presence of T-odd distribution functions and provide a deeper 
understanding of these phenomena.
\begin{figure}[t]
\vspace{-1cm}
\hspace{-0.55cm}
\mbox{~\epsfig{file=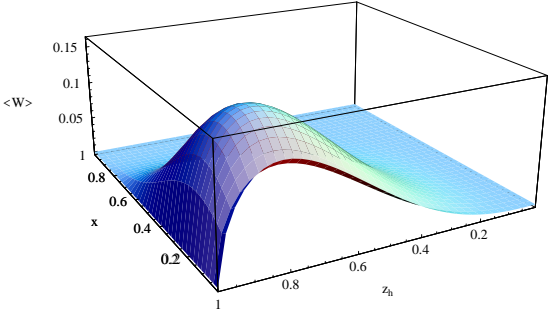,angle=0,width=7.5cm}}
\vspace{-1cm}
\caption{A~three-dimensional~view~of~the~quantity 
$\sum _{a,\overline a} e_a ^2 \ x \, h_{1}^{a}(x) \, 
H_1 ^{\perp (1) a} (z_h)$, for $OTO$ scattering with production of 
$\pi ^+$.}
\label{W3}
\vspace{-0.3cm}
\end{figure}
If instead  we choose the weight 
$W=(Q_T/M)\;\sin(\phi^l_h + \phi^l_S)$, we obtain an 
object which is directly proportional to the T-odd fragmentation function
$H_1^{\perp (1)}$ (see Table II, second line, in Ref.~\cite{bm98})
\bea
\lefteqn{
\left< \frac{Q_T}{M}\;\sin(\phi ^l _h + \phi ^l _S) \right>_{OTO} =}
\nonumber \\ && \hspace{-0.7cm}
\frac{ 4 \pi \alpha ^2 s}{Q^4} \, (1-y)\,
\sum _{a,\overline a} e_a ^2 \ x \, h_{1}^{a}(x)  \, H_1 ^{\perp (1) a} (z_h) 
\;.
\label{hH1}
\eea
As it clearly appears from the plot in Fig.~\ref{W3}, this time the 
shape of the quantity 
$\sum _{a,\overline a} e_a ^2 x \, h_{1}^{a}(x)  \, H_1 ^{\perp (1) a} (z_h)$ 
as a function of $x$ and $z_h$ is completely 
different from the previous one. It reaches its maximum 
for relatively small values of $x$ and for large values of $z_h$ and its 
overall size is at least a factor two bigger than the previous one. This 
means that a measure to reveal the effects of a non zero T-odd fragmentation 
function could easily be made at large values of $z_h$, where it is relatively 
easier to achieve  larger statistics.

Finally, we give an  evaluation of the ratios 
$H_1 ^{\perp a}  /  D_1 ^a$  and $f_{1T} ^{\perp} /  f_1$ (for 
$\pi ^+$ production and  considering only valence contributions). We find
\be
\left| \frac{\int_{0.1}^1 \, dz_h\ H_1^{\perp \mbox{\small fav}}(z_h)}
{\int_{0.1}^1 \, dz_h\ D_1^{u / \pi ^+} (z_h)} \right| = 0.076\;.
\ee
which gives  a value of about $8\%$, in agreement with the  
result of Ref.~\cite{est98}. For the T-odd distribution functions we have
\be
\left| \frac{\int _{0.02} ^{0.4} \, dx \, 
f_{1T}^{\perp \, u}(x)}{\int _{0.02} ^{0.4} \, dx \,f_1^{u} (x)} \right|
= 0.083\;,
\ee
\be
\left| \frac{\int _{0.02} ^{0.4} \, dx \, 
f_{1T}^{\perp \, d}(x)}{\int _{0.02} ^{0.4} \, dx \,f_1^{d} (x)} \right|
= 0.072\;,
\ee
which again gives an estimate of about 8\%.
We point out that the above estimates do not take into account the  effects of 
evolution and that comparing integrated results neglects some kinematics 
factors.

Another example is the $\sin\phi$ single spin asymmetry,
presented by the HERMES collaboration (see Avakian's contribution in 
these proceedings), corresponding to: 
\bea
\lefteqn{
\left< \frac{Q_T}{M}\;\sin(\phi ^l _h) \right>_{OLO} =
\frac{ 4 \pi \alpha ^2 s}{Q^4} \, (2-y)\,\sqrt(1-y)\,
} \nonumber \\ && \hspace{-0.7cm} 
\sum _{a,\overline a} e_a ^2 [ x 
 h_{1L}^{\perp (1) a}(x)   
\tilde{H} ^{a} (z_h) - x^2 h_{L}^{a}(x)  H_1 ^{\perp (1) a} (z_h)].
\nonumber
\eea
We are now able to give some estimates of this quantity, under suitable 
approximations: our calculation will be presented in a forthcoming 
paper~\cite{bm-2-99}.

\vspace{0.3cm}
\noindent
We acknowledge the support of the TMR program ERB FMRX-CT96-0008

\end{document}